\documentclass[twocolumn,showpacs,preprintnumbers,amsmath,amssymb]{revtex4}

\usepackage{graphicx}
\usepackage{dcolumn}
\usepackage{bm}
\usepackage{SIunits} 

\begin{document}

\preprint{For Applied Physics Letters}

\title{Opposite effects of NO$_2$ on electrical injection\\
in porous silicon gas sensors}

\author{Zeno Gaburro}
\email{gaburro@science.unitn.it}
\homepage{http://www.science.unitn.it/~semicon}
\author{Claudio J. Oton}
\author{Lorenzo Pavesi}

\affiliation{INFM and Department of Physics,
             University of Trento, Italy}

\author{Lucio Pancheri}
\affiliation{Department of Information and Communication
             Technology, University of Trento, Italy}

\date{\today}

\begin{abstract}
The electrical conductance of porous silicon fabricated with
heavily doped p-type silicon is very sensitive to NO$_2$. A
concentration of 10 ppb can be detected by monitoring the current
injection at fixed voltage. However, we show that the sign of the
injection variations depends on the porous layer thickness. If the
thickness is sufficiently low -- of the order of few
\micro\meter{} -- the injection decreases instead of increasing.
We discuss the effect in terms of an already proposed twofold
action of NO$_2$, according to which the free carrier density
increases, and simultaneously the energy bands are bent at the
porous silicon surface.
\end{abstract}



\maketitle

\newpage

Porous silicon (PSi) is obtained by electrical anodization of
crystalline silicon (Si) substrates~\cite{2000Bisi}. The internal
surface versus volume ratio of PSi can reach several hundreds of
\meter\squared{} per \centi\meter\cubed, leading to strong
dependence of optical and electrical properties of PSi on the
environment. For this reason, PSi is an interesting material for
gas sensors.

Recently, several reports have been focused on PSi sensors
fabricated using heavily doped p-type Si
(p$^+$ PSi), in which the sensitive parameter is the electrical
conductance~\cite{2000Boarino,2001Baratto,%
2001Boarino,2001Timoshenko,2003Pancheri,2003Chiesa}. In absence of
gases, the electrical injection in sufficiently thick ($\simeq$
30~\micro\meter) p$^+$ PSi layers is very low. In fact, the
anodization leads to a porous structure almost depleted of mobile
charges, despite the concentration of boron dopants is essentially
unchanged by the anodization, remaining approximately as high as
in the starting p$^+$ substrate
($\simeq10^{19}$\centi\meter$^{-3}$)
~\cite{1998Polisski,1999Polisski}. The absence of mobile charges
is ascribed to hole trapping~\cite{2001Boarino}, but the actual
mechanism is still controversial~\cite{2003Chiesa}. In presence of
specific gases, the mobile carrier population changes, usually
leading to increases of conductance, as in the
case of NO$_2$~\cite{2000Boarino,2001Baratto,%
2001Boarino,2001Timoshenko,2003Pancheri,2003Chiesa} or
NH$_3$~\cite{2003Chiesa}.

\begin{figure}
  \includegraphics[scale=0.5,clip]{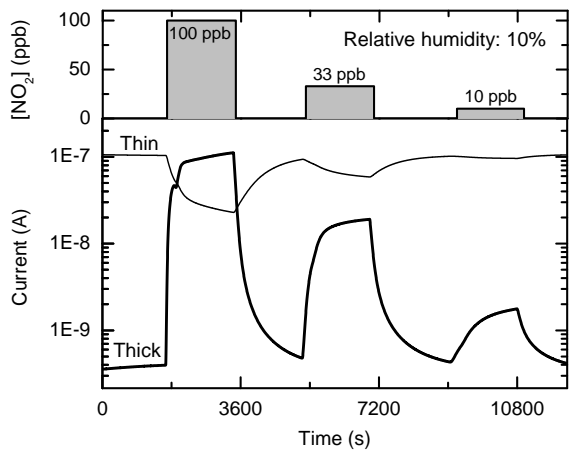}\\
  \caption{Simultaneous measurement of current injection in thick and thin
           sensors, under fixed bias voltage (bottom plot) and
           under controlled gas flux as a function of time. The bias
           voltage was -1~\volt{} applied to one of the top contacts
           with respect to the back contact, for both sensors. The
           top plot shows the composition of the gas flux.}\label{time}
\end{figure}

The presence of NO$_2$ is reported to free the trapped holes at
interface states~\cite{2001Boarino,2001Timoshenko}. As a
consequence, the conductivity increases because of the increased
free hole concentration~\cite{2000Boarino,2001Baratto,%
2001Boarino,2001Timoshenko,2003Pancheri}. The process is very
effective for sensing: the presence of NO$_2$ can be detected at
concentrations as low as 12~ppb~\cite{2003Pancheri}, and 10~ppb,
as shown below (Figure~\ref{time}). NO$_2$ is a product of
internal combustion engines and causes lung diseases. Most of
pollution regulations set the attention level of NO$_2$ around
100~ppb (for example, Italian Ministerial Decree, April 15, 1994).
Hence, the required level of sensitivity for realistic
applications is achievable in p$^+$~PSi.

The sensitivity to NO$_2$ is affected by both sensor's
porosity~\cite{2000Boarino} and micro
structure~\cite{2003Gaburro}. In this work, we report a quite
surprising effect: the electrical injection in p$^+$~PSi in
presence of NO$_2$ can increase or decrease, depending on the
thickness of the p$^+$~PSi layer. The known effect -- the increase
in the electrical injection originated by the increase of
conductance -- has been reported in \emph{thick} samples
($\simeq$ 30~\micro\meter)~\cite{2000Boarino,2001Baratto,%
2001Boarino,2001Timoshenko,2003Pancheri}. We report here for the
first time the opposite behavior of \emph{thin} samples ($\simeq$
2~\micro\meter), in which the injection \emph{decreases} in
presence of NO$_2$.

p$^+$~PSi layers were grown by electrochemical dissolution in an
HF-based solution on a single-crystalline p-type (100)
heavily-doped Si substrate. Substrate nominal resistivity $\rho$
was 6-15~\milli\ohm~\centi\meter. Before the anodization, the
native oxide was removed from the backside of the wafers, and
aluminium back contacts were deposited by evaporation. In order to
achieve high sensitivity to NO$_2$, we have used as anodizing
solution a mixture of 3 parts (in volume) of aqueous HF (48\% wt.)
and 7 parts of ethanol. The etching was performed by applying an
etching current density of
50~\milli\ampere\per\centi\meter\squared. We fabricated two types
of samples. The first type (hereafter referred to as ``thick'')
was fabricated using an etching time of 1363~\second, whereas the
second type (``thin'') was obtained with an etching time of
127~\second. Thus, thick and thin samples were fabricated with
exactly the same procedure, except for the anodization time. After
anodization, the samples were rinsed in ethanol and pentane, and
dried in ambient air. Scanning Electron Microscopy (SEM) images
showed a layer thickness of about 32 and 2~\micro\meter,
respectively, for thick and thin samples. From normal reflectance
measurements, a refractive index of about 1.4 was calculated, and
using Bruggeman approximation~\cite{2000Bisi} we have estimated
that porosity was about 78\%.

Gold electrodes were deposited by evaporation on the PSi top
surface. Copper wires were connected to the gold electrodes using
an epoxy silver paste.

\begin{figure}
  \includegraphics[scale=0.7]{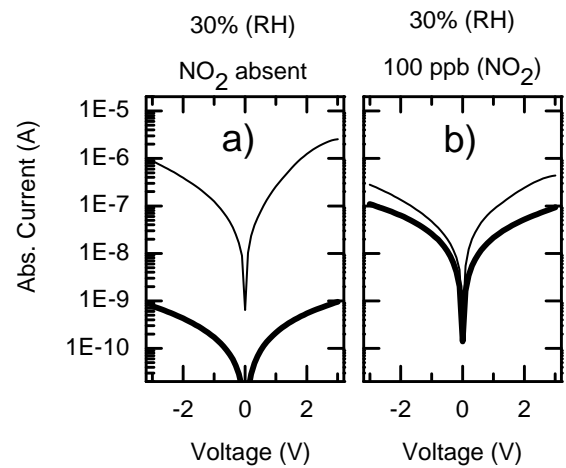}\\
  \caption{DC electrical characteristics of thick (thick
           line) and thin (thin line) sensors. The voltage sign is
           positive on the top contact with respect to the back.
           The composition
           of the gas flux is reported on top of each plot (RH=relative
           humidity).}\label{IV}
\end{figure}

For the electrical characterization, the sensors were kept in a
sealed chamber under controlled flux of gases coming from
certified cylinders. Humid air was obtained by flowing dry air
through a bubbler. Different relative humidity levels and NO$_2$
concentrations were obtained mixing humid air, dry air and a
dilute solution of NO$_2$ in air (550~ppb) with a flow control
system. Relative humidity was monitored using a calibrated
hygrometer. In order to characterize the time response, a bias
voltage of -1~\volt{} was applied on one of the top contacts with
respect to the back contact, and the injected current was
constantly monitored (Figure~\ref{time}).
Although the injected current does not totally stabilize after 30
minutes, data of Figure~\ref{time} suggest that the most
significant current variations are observed within such time
period. Thus, I-V characterization, performed as a separate
measurement, was acquired waiting an assumed 1800~\second{}
settling time after the gas switch. I-V characterizations are
reported in Figure~\ref{IV}.

The main result discussed in this work is clearly visible in both
Figures~\ref{time} and~\ref{IV}: in presence of NO$_2$, the
injected current increases in the thick sensor, and decreases in
the thin one.

In the case of the thick sensor, the impedance of the device is
dominated by the high resistance of the thick central layer.
During the exposure to NO$_2$, the conductance of the p$^+$~PSi
layer increases, as a consequence of the release of trapped holes,
as already established~\cite{2001Timoshenko,2001Boarino}. Thus,
the injection increases in presence of NO$_2$. In Figure~\ref{IV},
the increase is visible by comparing the thick lines of plot a)
and b). The I-V characteristic is symmetric with respect to a
change of sign in the voltage.

The behavior of the thin sensor is harder to explain. The first
observation is that in absence of NO$_2$ the current injection is
three orders of magnitude larger than in the thick sensor
(Figure~\ref{IV}, plot a), even though the thickness is only one
order of magnitude lower. This result suggests that in the thin
sensor, the resistivity of the porous layer does not significantly
limit the injection, which is therefore dominated by the behavior
of the two junctions.

The value of the work function of gold is similar to the work
function of p$^+$~Si ($\phi_{Au}\simeq5.1$~eV). Therefore, we
assume that the two junctions behave similarly, in terms of the
balance of excess charge at the junction regions. Since the
p$^+$~PSi is essentially an intrinsic
semiconductor~\cite{1998Polisski,1999Polisski}, the DC electrical
behavior should be similar to the one of a quasi-symmetric
\mbox{p$^+$-i-p$^+$} device. Although this is clearly a rough
simplification, nevertheless it seems a reasonable starting point,
since the I-V characteristics of the thin sensor are not very
asymmetric with respect to the voltage inversion
(Figure~\ref{IV}).

A recent study on \mbox{n$^+$-i-n$^+$} Si
structures~\cite{2000Cech} has shown that, in structures with
doping profiles similar to our thin sensor, when the distance
between the junctions is in the range of 2~\micro\meter{} or less,
extra free carriers diffusing from the heavily doped regions
accumulate over most of the intrinsic region. This result suggests
that the hole density in the porous layer of our thin sensor is
likely larger than the equilibrium density in bulk p$^+$~PSi,
because of the nearby junctions. In this sense, the two junctions
are close to each other. Also, it is reasonable to assume that
electron current density is negligible~\cite{2000Cech}. Finally,
Ref.~\cite{2000Cech} shows that in these structures the injection
can increase if the density of defect states is rearranged in
energy. We think that an energetic rearrangement at the surface is
the only way to reconcile an increase of free carrier density with
a decrease of the injected current, and to explain our
experimental results with the thin sensor.

\begin{figure}
  \includegraphics[scale=0.4]{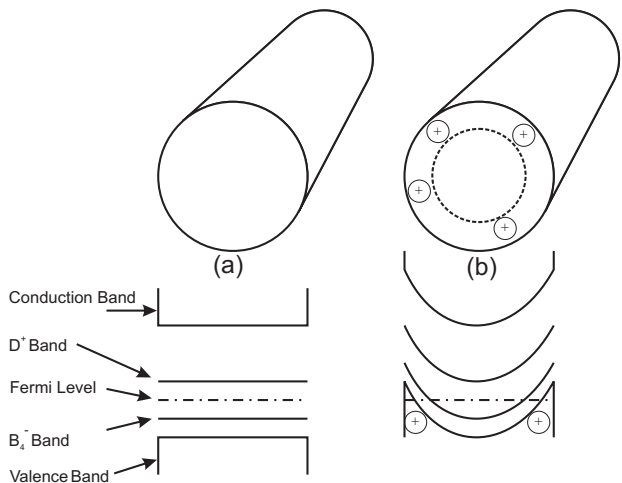}\\
  \caption{Pictorial representations of a conductive path of
           \emph{bulk} porous Si (thick sensor),
           in absence (a) and in presence (b) of
           NO$_2$. The energy states of each case are qualitatively
           shown
           underneath. In presence of NO$_2$, the effective
           cross-section of the conductive path is reduced by
           the band bending. In the thin sensor, free holes
           are present also in the case (a), and their density
           is homogeneous across the wire section, since the
           bands are flat.}\label{pillar}
\end{figure}

An interesting possibility is the detailed mechanism of
reactivation of boron dopants by NO$_2$ suggested by Boarino et
al.~\cite{2001Boarino}: in p$^+$~PSi, close the midgap energy,
there is a high density of dangling bonds ($D^0$). In absence of
gases, ionization of boron dopants can involve such defect states
($B_3^0+D^0\rightarrow B_4^-+D^+$). The process does not produce
free holes because of the trapping action of the defect states
($D$). The Fermi level remains between the energy levels of boron
and defect states, as schematized in Figure~\ref{pillar}, (a).
The presence of NO$_2$ bends the energy bands upward at the PSi
surface, pulling the Fermi level below the boron energy, and
restoring the original doping action of boron
(Figure~\ref{pillar},b)~\cite{2001Boarino}. An interesting feature
of this model is the energetically favorable location of holes at
the outer surface of conductive paths. In other words, the effect
of NO$_2$ can be envisioned as twofold: holes are de-trapped and
pulled at the surface. The latter aspect can be pictured as a
reduction of the effective conductive area (as in
Figure~\ref{pillar}, b) or more appropriately as a reduction of
hole mobility, because of larger surface scattering. If the hole
density is very low (as in the thick sensor), in absence of NO$_2$
there is hardly any conduction. The hole repopulation induced by
NO$_2$ totally obscures any effect on mobility. On the contrary,
if hole density is not negligible even in absence of NO$_2$, a
mobility reduction might be the only surviving effect of NO$_2$.
In the thin sensor, the hole density is higher than its
equilibrium value because of diffusion from the junctions. The
hole de-trapping by NO$_2$ is giving negligible contribution to
the overall hole density. Thus, we suppose that in thin sensors
the only significant effect is the the band bending, which reduces
the effective mobility, increasing the resistance.

If the bias is reversed, the situation is essentially the same,
except that the role of the two junctions is exchanged. The slight
asymmetry of I-V curves of thin sensors (Figure~\ref{IV}) is
probably due to this material asymmetry.

Finally, we wish to emphasize that the observed effect explains
why all the previous reports of p$^+$~PSi were focused on thick
sensors: it is not just a matter of increasing the sensitivity,
but rather of obscuring the contrasting effect of mobility
lowering, isolated here in thin sensors. In thick sensors, the
injection is dominated by free carrier repopulation.

In conclusion, in thin porous silicon sensors (thickness of the
order of few \micro\meter{} or less), the carrier density is
determined by the hole diffusion from the junctions. The surviving
effect of NO$_2$ in thin sensors is an effective narrowing of the
conductive cross-sections, which in turn lowers the effective hole
mobility. As a consequence, the net effect of NO$_2$ in thick and
thin porous silicon sensors is opposite in sign. For this reason,
the gas sensing effects related to modulation of free carrier
concentration are best exploited if the porous layer thickness is
at least a few tens of \micro\meter.

We acknowledge the support of INFM, progetto PAIS 2001 "SMOG" and
of Provincia Autonoma di Trento.

\bibliography{2003_thinthick}

\begin{thebibliography}{11}
\expandafter\ifx\csname natexlab\endcsname\relax\def\natexlab#1{#1}\fi
\expandafter\ifx\csname bibnamefont\endcsname\relax
  \def\bibnamefont#1{#1}\fi
\expandafter\ifx\csname bibfnamefont\endcsname\relax
  \def\bibfnamefont#1{#1}\fi
\expandafter\ifx\csname citenamefont\endcsname\relax
  \def\citenamefont#1{#1}\fi
\expandafter\ifx\csname url\endcsname\relax
  \def\url#1{\texttt{#1}}\fi
\expandafter\ifx\csname urlprefix\endcsname\relax\def\urlprefix{URL }\fi
\providecommand{\bibinfo}[2]{#2}
\providecommand{\eprint}[2][]{\url{#2}}

\bibitem[{\citenamefont{Bisi et~al.}(2000)\citenamefont{Bisi, Ossicini, and
  Pavesi}}]{2000Bisi}
\bibinfo{author}{\bibfnamefont{O.}~\bibnamefont{Bisi}},
  \bibinfo{author}{\bibfnamefont{S.}~\bibnamefont{Ossicini}}, \bibnamefont{and}
  \bibinfo{author}{\bibfnamefont{L.}~\bibnamefont{Pavesi}},
  \bibinfo{journal}{Surface Science Reports} \textbf{\bibinfo{volume}{38}},
  \bibinfo{pages}{1} (\bibinfo{year}{2000}).

\bibitem[{\citenamefont{Boarino et~al.}(2000)\citenamefont{Boarino, Baratto,
  Geobaldo, Amato, Comini, Rossi, Faglia, L\'erondel, and
  Sberveglieri}}]{2000Boarino}
\bibinfo{author}{\bibfnamefont{L.}~\bibnamefont{Boarino}},
  \bibinfo{author}{\bibfnamefont{C.}~\bibnamefont{Baratto}},
  \bibinfo{author}{\bibfnamefont{F.}~\bibnamefont{Geobaldo}},
  \bibinfo{author}{\bibfnamefont{G.}~\bibnamefont{Amato}},
  \bibinfo{author}{\bibfnamefont{E.}~\bibnamefont{Comini}},
  \bibinfo{author}{\bibfnamefont{A.~M.} \bibnamefont{Rossi}},
  \bibinfo{author}{\bibfnamefont{G.}~\bibnamefont{Faglia}},
  \bibinfo{author}{\bibfnamefont{G.}~\bibnamefont{L\'erondel}},
  \bibnamefont{and}
  \bibinfo{author}{\bibfnamefont{G.}~\bibnamefont{Sberveglieri}},
  \bibinfo{journal}{Materials Science and Engineering B}
  \textbf{\bibinfo{volume}{69-70}}, \bibinfo{pages}{210}
  (\bibinfo{year}{2000}).

\bibitem[{\citenamefont{Baratto et~al.}(2001)\citenamefont{Baratto, Faglia,
  Sberveglieri, Boarino, Rossi, and Amato}}]{2001Baratto}
\bibinfo{author}{\bibfnamefont{C.}~\bibnamefont{Baratto}},
  \bibinfo{author}{\bibfnamefont{G.}~\bibnamefont{Faglia}},
  \bibinfo{author}{\bibfnamefont{G.}~\bibnamefont{Sberveglieri}},
  \bibinfo{author}{\bibfnamefont{L.}~\bibnamefont{Boarino}},
  \bibinfo{author}{\bibfnamefont{A.~M.} \bibnamefont{Rossi}}, \bibnamefont{and}
  \bibinfo{author}{\bibfnamefont{G.}~\bibnamefont{Amato}},
  \bibinfo{journal}{Thin Solid Films} \textbf{\bibinfo{volume}{391}},
  \bibinfo{pages}{261} (\bibinfo{year}{2001}).

\bibitem[{\citenamefont{Boarino et~al.}(2001)\citenamefont{Boarino, Geobaldo,
  Borini, Rossi, Rivolo, Rocchia, Garrone, and Amato}}]{2001Boarino}
\bibinfo{author}{\bibfnamefont{L.}~\bibnamefont{Boarino}},
  \bibinfo{author}{\bibfnamefont{F.}~\bibnamefont{Geobaldo}},
  \bibinfo{author}{\bibfnamefont{S.}~\bibnamefont{Borini}},
  \bibinfo{author}{\bibfnamefont{A.~M.} \bibnamefont{Rossi}},
  \bibinfo{author}{\bibfnamefont{P.}~\bibnamefont{Rivolo}},
  \bibinfo{author}{\bibfnamefont{M.}~\bibnamefont{Rocchia}},
  \bibinfo{author}{\bibfnamefont{E.}~\bibnamefont{Garrone}}, \bibnamefont{and}
  \bibinfo{author}{\bibfnamefont{G.}~\bibnamefont{Amato}},
  \bibinfo{journal}{Physical Review B} \textbf{\bibinfo{volume}{64}},
  \bibinfo{pages}{205308} (\bibinfo{year}{2001}).

\bibitem[{\citenamefont{Timoshenko et~al.}(2001)\citenamefont{Timoshenko,
  Dittrich, Lysenko, Lisachenko, and Koch}}]{2001Timoshenko}
\bibinfo{author}{\bibfnamefont{V.~Y.} \bibnamefont{Timoshenko}},
  \bibinfo{author}{\bibfnamefont{T.}~\bibnamefont{Dittrich}},
  \bibinfo{author}{\bibfnamefont{V.}~\bibnamefont{Lysenko}},
  \bibinfo{author}{\bibfnamefont{M.~G.} \bibnamefont{Lisachenko}},
  \bibnamefont{and} \bibinfo{author}{\bibfnamefont{F.}~\bibnamefont{Koch}},
  \bibinfo{journal}{Physical Review B} \textbf{\bibinfo{volume}{64}},
  \bibinfo{pages}{085314} (\bibinfo{year}{2001}).

\bibitem[{\citenamefont{Pancheri et~al.}(2003)\citenamefont{Pancheri, Ot\'on,
  Gaburro, Soncini, and Pavesi}}]{2003Pancheri}
\bibinfo{author}{\bibfnamefont{L.}~\bibnamefont{Pancheri}},
  \bibinfo{author}{\bibfnamefont{C.~J.} \bibnamefont{Ot\'on}},
  \bibinfo{author}{\bibfnamefont{Z.}~\bibnamefont{Gaburro}},
  \bibinfo{author}{\bibfnamefont{G.}~\bibnamefont{Soncini}}, \bibnamefont{and}
  \bibinfo{author}{\bibfnamefont{L.}~\bibnamefont{Pavesi}},
  \bibinfo{journal}{Sensors and Actuators B} \textbf{\bibinfo{volume}{89}},
  \bibinfo{pages}{237} (\bibinfo{year}{2003}).

\bibitem[{\citenamefont{Chiesa et~al.}(2003)\citenamefont{Chiesa, Amato,
  Boarino, Garrone, Geobaldo, and Giamello}}]{2003Chiesa}
\bibinfo{author}{\bibfnamefont{M.}~\bibnamefont{Chiesa}},
  \bibinfo{author}{\bibfnamefont{G.}~\bibnamefont{Amato}},
  \bibinfo{author}{\bibfnamefont{L.}~\bibnamefont{Boarino}},
  \bibinfo{author}{\bibfnamefont{E.}~\bibnamefont{Garrone}},
  \bibinfo{author}{\bibfnamefont{F.}~\bibnamefont{Geobaldo}}, \bibnamefont{and}
  \bibinfo{author}{\bibfnamefont{E.}~\bibnamefont{Giamello}},
  \bibinfo{journal}{Angew. Chem. Int. Ed.} \textbf{\bibinfo{volume}{42}},
  \bibinfo{pages}{5032} (\bibinfo{year}{2003}).

\bibitem[{\citenamefont{Polisski et~al.}(1998)\citenamefont{Polisski,
  Dollinger, Bergmaier, Kovalev, Heckler, and Koch}}]{1998Polisski}
\bibinfo{author}{\bibfnamefont{G.}~\bibnamefont{Polisski}},
  \bibinfo{author}{\bibfnamefont{G.}~\bibnamefont{Dollinger}},
  \bibinfo{author}{\bibfnamefont{A.}~\bibnamefont{Bergmaier}},
  \bibinfo{author}{\bibfnamefont{D.}~\bibnamefont{Kovalev}},
  \bibinfo{author}{\bibfnamefont{H.}~\bibnamefont{Heckler}}, \bibnamefont{and}
  \bibinfo{author}{\bibfnamefont{F.}~\bibnamefont{Koch}},
  \bibinfo{journal}{phys. stat. sol. (a)} \textbf{\bibinfo{volume}{168}},
  \bibinfo{pages}{R1} (\bibinfo{year}{1998}).

\bibitem[{\citenamefont{Polisski et~al.}(1999)\citenamefont{Polisski, Kovalev,
  Dollinger, Sulima, and Koch}}]{1999Polisski}
\bibinfo{author}{\bibfnamefont{G.}~\bibnamefont{Polisski}},
  \bibinfo{author}{\bibfnamefont{D.}~\bibnamefont{Kovalev}},
  \bibinfo{author}{\bibfnamefont{G.}~\bibnamefont{Dollinger}},
  \bibinfo{author}{\bibfnamefont{T.}~\bibnamefont{Sulima}}, \bibnamefont{and}
  \bibinfo{author}{\bibfnamefont{F.}~\bibnamefont{Koch}},
  \bibinfo{journal}{Physica B} \textbf{\bibinfo{volume}{263-274}},
  \bibinfo{pages}{951} (\bibinfo{year}{1999}).

\bibitem[{\citenamefont{Gaburro et~al.}(2003)\citenamefont{Gaburro, Bettotti,
  Saiani, Pavesi, Pancheri, Oton, and Capuj}}]{2003Gaburro}
\bibinfo{author}{\bibfnamefont{Z.}~\bibnamefont{Gaburro}},
  \bibinfo{author}{\bibfnamefont{P.}~\bibnamefont{Bettotti}},
  \bibinfo{author}{\bibfnamefont{M.}~\bibnamefont{Saiani}},
  \bibinfo{author}{\bibfnamefont{L.}~\bibnamefont{Pavesi}},
  \bibinfo{author}{\bibfnamefont{L.}~\bibnamefont{Pancheri}},
  \bibinfo{author}{\bibfnamefont{C.~J.} \bibnamefont{Oton}}, \bibnamefont{and}
  \bibinfo{author}{\bibfnamefont{N.}~\bibnamefont{Capuj}},
  \bibinfo{journal}{sumbitted to Appl. Phys Lett.}  (\bibinfo{year}{2003}).

\bibitem[{\citenamefont{Cech}(2000)}]{2000Cech}
\bibinfo{author}{\bibfnamefont{V.}~\bibnamefont{Cech}},
  \bibinfo{journal}{Journal of Appplied Physics} \textbf{\bibinfo{volume}{88}},
  \bibinfo{pages}{5374} (\bibinfo{year}{2000}).

\end{thebibliography}

\end{document}